\documentclass[%
reprint,
superscriptaddress,
nobibnotes,
amsmath,amssymb,
aps,
prb,
]{revtex4-1}

\usepackage{graphicx}% Include figure files
\usepackage{dcolumn}% Align table columns on decimal point
\usepackage{bm}% bold math
\usepackage{hyperref}% add hypertext capabilities
\hypersetup{
    breaklinks=true,
    bookmarks=true,
    pdfauthor={},
    pdftitle={},
    colorlinks=true,
    citecolor=blue,
    urlcolor=blue,
    linkcolor=magenta,
    pdfborder={0 0 0}
}

\usepackage{mathtools} % For pmatrix* s.t. can adjust alignment
\usepackage{upgreek} % For upright (non-italic) lowercase greek symbols
\usepackage{color,soul} %For highlighting
\usepackage[normalem]{ulem} %For underline formatting
\usepackage{xfrac} %To use slanted fractions
\usepackage{braket} %To use braket notations
\usepackage{cases} %To use numbered cases

\def\abinit{\textit{ab initio} }
\newcommand{\kfPlusGPar}{\mathbf{k}_{f \parallel} + \mathbf{G}_\parallel}

\begin{document}

\preprint{APS/123-QED}

\title{\textit{Ab initio} many-body photoemission theory of transverse energy distribution of photoelectrons: PbTe(111) as a case study with experimental comparisons}

% Force line breaks with \\

\author{J. Kevin Nangoi}
%\altaffiliation[Also at ]{%Physics Department, XYZ University. Department of
%Physics, Cornell University, Ithaca, NY 14850}
%Lines break automatically or can be forced with \\ \author{Second Author}%
\email[Corresponding author. Email: ]{jn459@cornell.edu}
\affiliation{
    Department of Physics, Cornell University, Ithaca, New York 14853, USA
}

\author{Siddharth Karkare}
\altaffiliation[Present address: ]{
    Department of Physics, Arizona State University, Tempe, Arizona 85287, USA
}
\affiliation{
    Lawrence Berkeley National Laboratory, 1 Cyclotron Rd., Berkeley, California 94720, USA
}

\author{Ravishankar Sundararaman}
\affiliation{
    Department of Materials Science and Engineering, Rensselaer Polytechnic Institute, Troy, New York 12180, USA
}

\author{Howard A. Padmore}
\affiliation{
    Lawrence Berkeley National Laboratory, 1 Cyclotron Rd., Berkeley, California 94720, USA
}

\author{Tom\'{a}s A. Arias}
\affiliation{
    Department of Physics, Cornell University, Ithaca, New York 14853, USA
}


%\affiliation{
%\affiliation{
%\affiliation{%

%             %  but any date may be explicitly specified

\begin{abstract}

    This manuscript presents, to our knowledge, the first fully \abinit many-body photoemission framework to predict the transverse momentum distributions and the mean transverse energies (MTEs) of photoelectrons from single-crystal photocathodes. 
    The need to develop such a theory stems from the lack of studies that provide complete understanding of the underlying fundamental processes governing the transverse momentum distribution of photoelectrons emitted from single crystals. 
    For example, initial predictions based on density-functional theory calculations of effective electron masses suggested that the (111) surface of PbTe would produce very small MTEs ($\leq$ 15 meV), whereas our experiments yielded MTEs {\em ten to twenty} times larger than these predictions, and also exhibited a lower photoemission threshold than predicted.
    The \abinit framework presented in this manuscript correctly reproduces the magnitude of the MTEs from our measurements in PbTe(111) and also the observed photoemission below the predicted threshold.
    Our results show that photoexcitations into bulk-like states and coherent, many-body electron-photon-phonon scattering processes, both of which initial predictions ignored, indeed play important roles in photoemission from PbTe(111). 
    Finally, from the lessons learned, we recommend a procedure for rapid computational screening of potential single-crystal photocathodes for applications in next-generation 
    ultrafast electron diffraction and X-ray free-electron lasers, which will enable new, significant advances in condensed matter research.

\end{abstract}

%\begin{description}
%You may use the \texttt{description} environment to structure your %abstract;
%use the optional argument of the \verb+\item+ command to give the %category of
%each item. 
%\end{description}

\maketitle

%\tableofcontents

%\section{\label{sec:level1}First-level heading} default

\section{Introduction} \label{sec:intro}

Mean transverse energy (MTE), the average kinetic energy of photoemitted
electrons \textit{parallel} to a photocathode surface, is a key quantity that
limits the brightness of state-of-the-art laser-driven electron
sources~\cite{ref:cathode-RnD, ref:ag-111_model} used, for example, 
in ultrafast electron diffraction (UED)~\cite{ref:uedm-4d}
and X-ray free-electron lasers (XFELs).\cite{ref:xfel, ref:cathode-RnD}
Reducing the MTE increases the electron beam brightness, which increases the
spatial resolution of UED\cite{ref:rh-110, ref:rh-110_ref8, ref:rh-110_ref11,
ref:ag-111, ref:ag-111_ref7} and the maximum lasing photon energy of
XFELs.\cite{ref:rh-110, ref:rh-110_ref10, ref:ag-111, ref:ag-111_ref6} 
Increased electron beam brightness also will enable more thorough and accurate
studies of various physical phenomena, including the ultrafast photo-induced
metal-insulator transition of VO$_2$\cite{ref:VO2} and ultrafast
photo-conversion dynamics in rhodopsin.\cite{ref:rhodopsin}

In the absense of a comprehensive theory, early efforts to reduce the mean transverse energy involve operating
photocathodes both at cryogenic temperatures and near the photoemission
threshold.\cite{ref:mte_thermal-limit, ref:alk-ant_cryo, ref:adv-bri-src-2018} 
However, these conditions result in very low quantum efficiency, the number of emitted photoelectrons
per incident laser photon, and thus can lower the overall beam brightness despite the
reduced MTE.\cite{ref:adv-bri-src-2018} 
To address this limitation, more recent efforts to lower the MTE without sacrificing quantum
efficiency focus on \textit{single-crystal} photocathodes, which possess well-defined band structures that can be exploited to produce low-MTE electron beams.\cite{ref:adv-bri-src-2018}
\citet{ref:ag-111} presented one of the first successful experimental attempts to
reduce the MTE using single-crystal photocathodes. They found that single-crystal Ag(111) reduces MTEs below that of polycrystalline Ag\cite{ref:schroeder_metal-eff-mass} 
and yields a significantly larger quantum efficiency than a typical polycrystalline metal due to the high density-of-states close to the Fermi level provided by a Shockley surface state.\cite{ref:ag-111}

Despite the recent experimental progress with single-crystal photocathodes, there are to date no fully \abinit studies exploring the fundamental underlying
physics and predicting the resulting MTE. 
Early photocathode MTE theory focused primarily on polycrystals or photocathodes with disordered surfaces and used semi-empirical approaches.\cite{ref:FEL-report, ref:jensen, ref:dowell, ref:karkare_GaAs-MC,
ref:mte_thermal-limit, ref:alk-ant_cryo} 
\citet{ref:schroeder-one-step} do consider single-crystal photocathodes, but
calculate the MTE by considering only electrons that originate
from bulk states near the Fermi level and treating them as free electrons. 
\citet{ref:ag-111_model} developed a theoretical model that explains their MTE
measurements on Ag(111)~\cite{ref:ag-111} by using the nearly free-electron model, 
a good approximation for only a limited number of metals.\cite{ref:AnM_NFE}
\citet{ref:rh-110},~\citet{ref:li-sch-p3}, and~\citet{ref:li-sch} use the free-electron approximation informed with 
\abinit effective masses calculated from density-functional theory (DFT), but neglect the full band structure of the material.
Finally, \citet{ref:li-sch_bcc-metals} use the full DFT band structure, but
approximate the photoexcitation transition rates as uniform instead of
calculating them \abinit from the appropriate transition matrix elements.

Beyond not being fully \textit{ab initio}, most of the above single-crystal MTE studies assume direct photoexcitation into
vacuum states.
In reality, electrons also transition into propagating states that extend well into the bulk of the material. Such transitions can become the dominant process because, at typical operating laser energies, laser photons penetrate the many atomic layers beneath the surface and can excite large numbers of electrons from the bulk region.
For example, in PbTe, at the laser energies of interest ($\sim$4--5~eV), photons
have a characteristic absorption depth of $\sim$200
\AA,\cite{ref:pbte_absCoef} approximately thirty times the lattice
constant.\cite{ref:pbte_exp-latt-const}

Relatively recently, \citet{ref:camino_ab-initio-QE} did present a fully \abinit treatment of such bulk-like excitation processes in single crystals, but, due to a different focus, considered only quantum efficiency rather than the momentum and energy distributions of the emitted photoelectrons.  They do consider a full, {\em ab initio} band structure and 
transitions of bulk electrons into bulk-like propagating states, 
with the transition rates calculated appropriately from first principles using Fermi's golden rule. However, due to their focus on quantum efficiency as opposed to MTE,
\citet{ref:camino_ab-initio-QE} do not consider 
many-body excitation processes such as electron-phonon scattering, which can alter the momentum distribution of excited electrons.\cite{ref:karkare_GaAs-MC}

Finally, immediately before submission of this manuscript, we became aware of an in-production abstract of an even more recent work focusing directly on photoemission and mean transverse energy, as well as intrinsic emittance in PbTe(111) and other materials. \cite{ref:antoniuk} This last work considers bulk processes, but does not include many-body processes, such as the electron-photon-phonon excitation we find to be important below. Moreover, in addition to applying the well-established ``scissors operator,''\cite{ref:scissors1, ref:scissors2} Ref.~\onlinecite{ref:antoniuk} finds that, in order to match the MTE versus photon energy profile for PbTe(111), it is necessary within their framework to employ a novel ``stretching'' operator that scales the entire conduction band structure by a constant multiplicative factor. The authors of Ref.~\onlinecite{ref:antoniuk} choose a factor of $\sim$0.6 to match the effective mass at the conduction band minimum; however, as we demonstrate below (Figure~\ref{fig:bandstruct_trans-mom-distrib_dir}(a)), the lower conduction band states do not participate in the photoemission process. Moreover, such a ``stretch'' significantly distorts the energy scales of the remainder of the density-functional theory conduction band structure, particularly the energies of those transitions which actually do participate in photoemission. For example, this scaling reduces the primary transition from an energy of $\sim$4.5~eV to $\sim$2.8~eV. Given that the density-functional theory band structure, after a scissors operator and without ``stretching,'' agrees quite well with established many-body GW calculations,\cite{ref:pbte_gw} the agreement with experiment found in Ref.~\onlinecite{ref:antoniuk} does not reflect the underlying physics.

Further progress in understanding the fundamentals of photoelectron distributions and MTEs thus requires both development of a new \abinit framework that considers many-body processes such as electron-phonon scattering \textit{and} comparison with experiments exploring the full distributions of emitted photoelectrons.
This manuscript presents just such a theory applicable to any single-crystal
photocathode material, as well as comparisons to detailed experiments which
measure not only the MTE but also the full distribution of transverse electron momenta.
As a case study, we consider single-crystal, semiconducting PbTe(111), finding
good agreement with experiment, and we explain the significant discrepancies
between our experimental measurements and previous theoretical
estimates.\cite{ref:li-sch, ref:li-sch-p3}
We find that 
consideration of 
electronic transitions into bulk-like propagating states
and inclusion of coherent three-body electron-photon-phonon scattering 
to be key in explaining the observed 
MTEs, and develop significant insight into the underlying physical mechanisms.

%===============================================================
%                   END - INTRO
%===============================================================

\section{Previous Theories vs. Experiment} \label{sec:prev-vs-exp}

\begin{figure}[h!]
    \centering
    \includegraphics[width=0.9\linewidth]{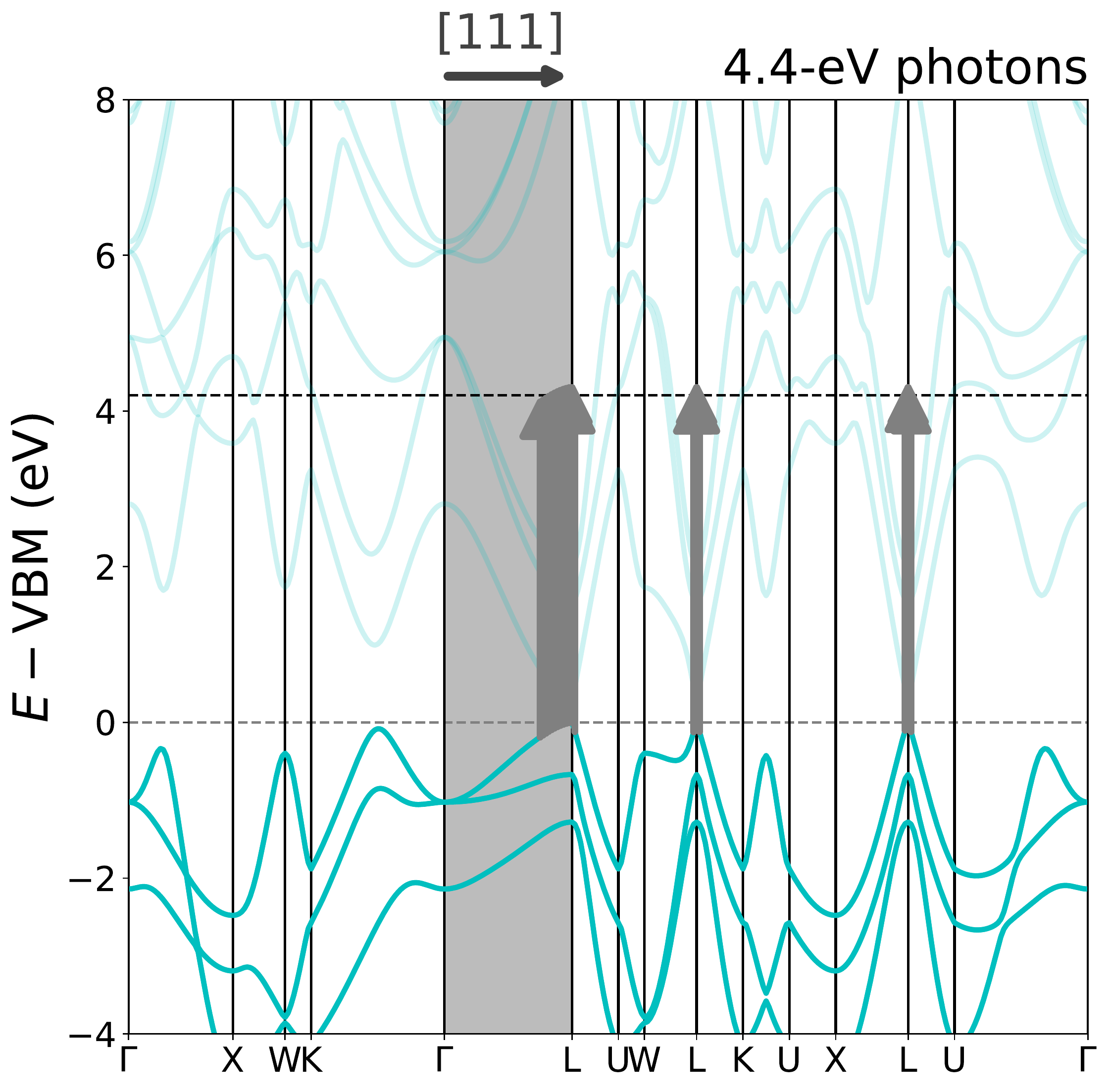}

    \caption{
        Bulk band structure of PbTe: valence bands (solid curves), conduction
        bands (faint solid curves), high-symmetry path $\Gamma$--L 
        parallel to the $[$111$]$ surface normal direction (shaded region), valence band maximum (horizontal dashed line at 0), vacuum level
        from Refs. \onlinecite{ref:li-sch-p3} and \onlinecite{ref:li-sch}
        (horizontal dashed line at 4.2~eV), and vertical transitions
        \textit{directly into vacuum states} considered in Refs.
        \onlinecite{ref:li-sch-p3} and \onlinecite{ref:li-sch} at a photon
        energy of 4.4~eV (vertical arrows). 
        Arrow thickness corresponds to the number of available transitions and is proportional to both the density-of-states and the effective mass.
    }
    
    \label{fig:bandstruct_sch} 
\end{figure}

Motivated by the need for mean transverse energies below 10 meV for next-generation ultra-high-brightness applications,\cite{ref:ultrabright} previous density-functional theory studies~\cite{ref:li-sch,
ref:li-sch-p3} identified single-crystal, semiconducting PbTe(111) as capable of producing
very low MTEs ($\leq 15$ meV). 
These studies attribute the low MTEs to the small transverse effective masses associated with the valence band maximum of the material, as well as the impact of these small masses on direct transitions into vacuum states. 
Figure~\ref{fig:bandstruct_sch} illustrates such transitions for a photon energy
of 4.4~eV, $\sim$0.2 eV above the calculated threshold from Refs.
\onlinecite{ref:li-sch-p3} and \onlinecite{ref:li-sch}. 
Due to the small transverse effective masses, there are very few allowed transitions with small, non-zero transverse momenta along the W--L, L--K, X--L, and L--U directions.
In contrast, there are a large number of transitions along $\Gamma$--L due to the larger effective mass along this direction. Because the $\Gamma$--L direction is parallel to the (111) surface normal, the corresponding photoelectrons will have essentially zero transverse momenta.
These observations are what ultimately lead Refs. \onlinecite{ref:li-sch-p3} and \onlinecite{ref:li-sch} to predict a very low MTE of $\leq$ 15 meV at this photon energy.

To explore the above predictions, we here measure the MTEs of photoelectrons emitted from an atomically ordered Pb-terminated PbTe(111) surface. We first prepared the surface by performing several cycles of ion-bombardment with 500 eV Ar$^+$ ions followed by annealing to 260$^\circ$C on a commercially-purchased single-crystal PbTe(111) substrate.\cite{ref:pbte111_recons_exp,ref:pbte_prep2} 
We continued these cycles until the surface exhibited a sharp 1$\times$1 hexagonal low-energy electron diffraction pattern and until Auger electron spectroscopy showed no surface contaminants. The light source generating the photoelectrons is a laser-based plasma lamp with a tunable wavelength monochromator\cite{ref:monochromator} and a spectral width of 2 nm FWHM. 
The incident light is 35$^\circ$ off-normal and has a focused spot diameter of $\sim$150 $\mu$m on the PbTe(111) surface.
We then accelerate the resulting photoelectron beam longitudinally to several kilovolts through a flat fine-mesh anode, allowing the beam to drift and expand from the effective point-spot on the PbTe(111) cathode under the transverse momenta of the photoelectrons. 
We obtain the transverse momentum distribution and MTE by measuring the size of the photoelectron beam after the acceleration with the setup given in Ref.~\onlinecite{ref:momentatron}. 

\begin{figure}[h!]
    \centering
    \includegraphics[width=0.9\linewidth]{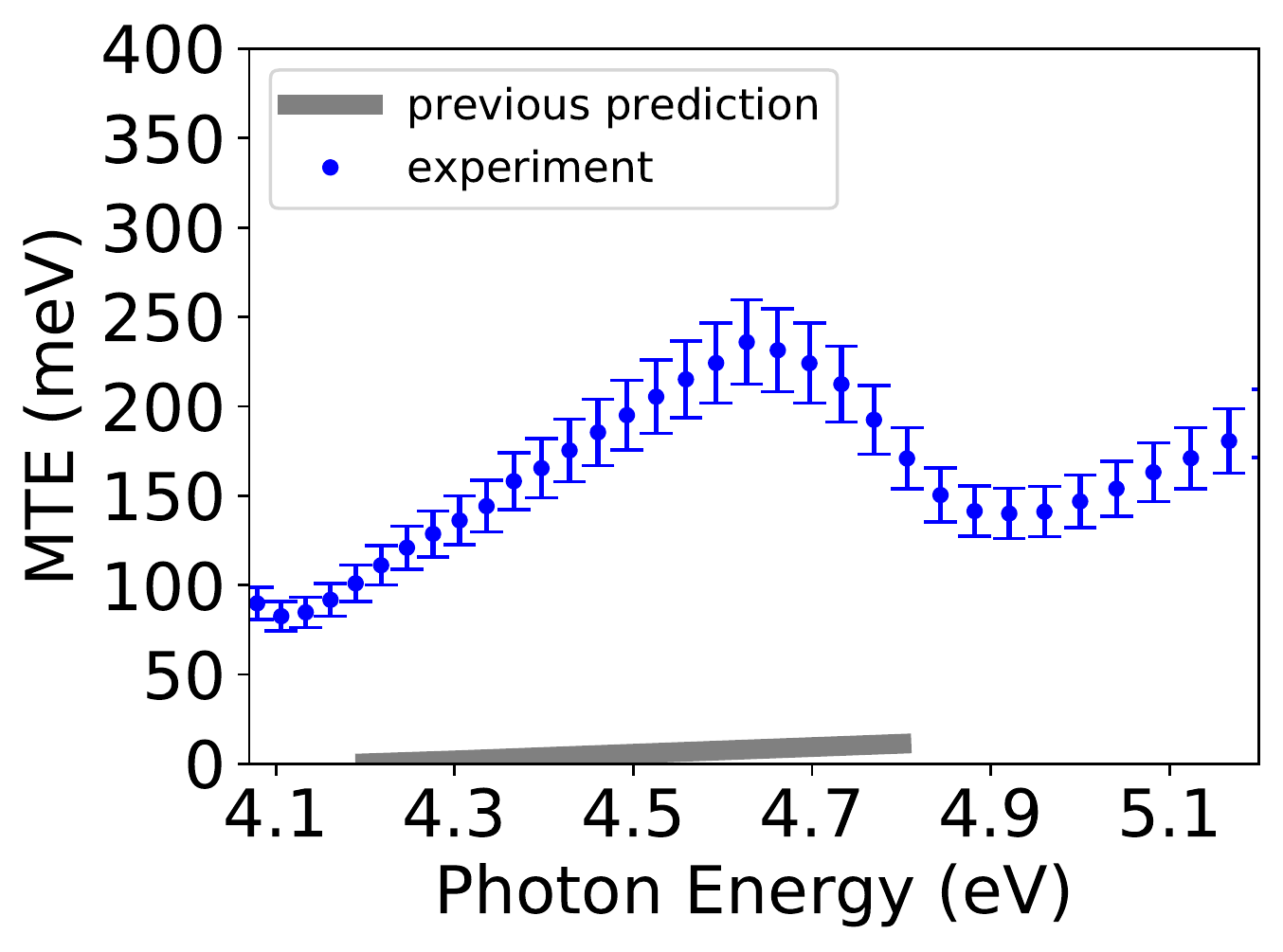}

    \caption{
        Mean transverse energy (MTE) of photoelectrons emitted from PbTe(111) as a function of laser photon energy: 
        our experimental results at room temperature (points with error bars) 
        and previous predictions\cite{ref:li-sch, ref:li-sch-p3} (thick curve), which are 10--20$\times$ smaller than the observed MTEs and exhibit a higher threshold energy of 4.2 eV.
    }
    \label{fig:MTE-init}
\end{figure}

Figure~\ref{fig:MTE-init} contrasts the MTE predictions from Refs. \onlinecite{ref:li-sch-p3} and
\onlinecite{ref:li-sch} with our experimentally-measured MTEs at room temperature. 
Our measured values are up to twenty times larger than predicted. 
Moreover, unlike the predictions, our measurements exhibit non-monotonic behavior as a
function of photon energy, as well as photoemission below the predicted
threshold of 4.2~eV.

The significant discrepancies in magnitude and trend between our experimental
MTEs and the predicted MTEs suggest additional processes to be at work.
For example, excitations into bulk-like propagating states that can transmit electrons into vacuum may impact significantly the allowed transitions and affect the final distribution of emitted photoelectrons.
Moreover, the observed photoemission below the predicted threshold suggests 
the presence of indirect photoexcitations, much like how indirect excitations
in semiconductors can occur below the direct band gap. 
Therefore, coherent three-body electron-photon-phonon scattering also may
play an important role in photoemission from PbTe(111). 
To explore these possibilities we now develop an \abinit framework capable of including these processes and predicting the momentum and energy distributions of the resulting photoelectrons.

%===============================================================
%                   END - Previous vs. Exp
%===============================================================

\section{Fully \textit{Ab Initio} Approach} \label{sec:methods}

To better explain our experimental observations of the MTE of PbTe(111), new
theory must be developed.  As observed in the previous section, such a theory
must not only account for direct photoexcitations into the bulk-like states
that can transmit electrons into vacuum, but also account for coherent
electron-photon-phonon scattering.  We previously reported in conference
proceedings the results of calculations accounting for the above
effects.\cite{ref:me_ipac} However, those calculations did not account for a
number of other important factors.  First, the finite linewidths of the
intermediate electron states during the electron-photon-phonon scattering
process can impact significantly the indirect photoexcitation transition rates
and thus must be included in the calculations. Second, the Bloch components of
an excited bulk-like state in the material couple to \textit{not merely} a
single outgoing plane-wave component in the vacuum, \textit{but to any}
plane-wave component whose momentum along the surface differs by any reciprocal
lattice vector of the two-dimensional surface lattice.  Third, the laser light
in the experiment may not be of a single polarization, but may be unpolarized,
as in our experiments. Finally, as is well known, band gaps calculated using
density-functional theory can be inaccurate and should be corrected through the
standard so-called ``scissors operator'',\cite{ref:scissors1, ref:scissors2} a
procedure we had not yet applied.

The following subsections lay out our final method for calculating photoemission, transverse momentum distributions, and MTEs. 
First, Subsection~\ref{subsec:framework} describes the overall framework which we use to 
address all of the above issues
and to calculate the transverse momentum distributions and MTEs.
Subsections 
\ref{subsec:rates}, 
\ref{subsec:surfTrans}, 
and \ref{subsec:details} then 
give details, respectively, 
of how we compute the photoexcitation transition rates, 
of how we calculate the probabilities of transmission into vacuum, 
and of the specific computational aspects of the underlying calculations.

%---------------------------------------------------------
%          Methods: Overall Framework
%---------------------------------------------------------
\subsection{Overall Framework} \label{subsec:framework}

As described in previous sections, this work considers photoelectrons that originate from bound bulk-like states of the photocathode material
and then transition into higher-energy bulk-like states that propagate in vacuum. 
This subsection describes our general photoemission framework, which is applicable to single-crystal metallic as well as semiconducting photocathodes. 

\begin{figure}[h!]
    \centering
    \includegraphics[width=0.9\linewidth]{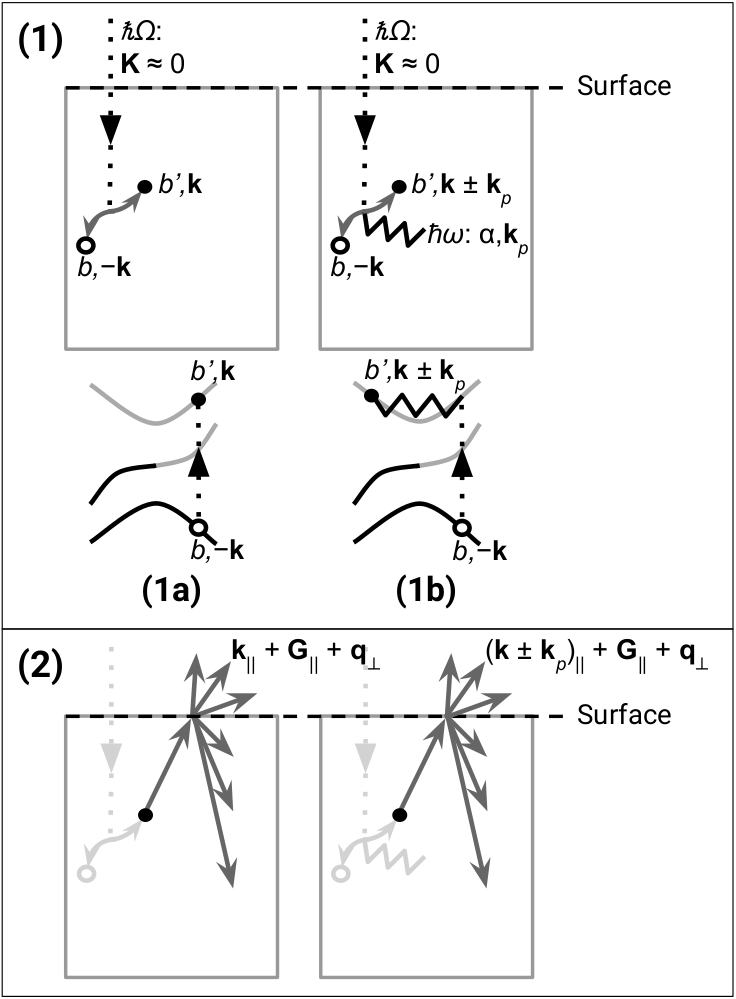}

    \caption{
        Steps of the many-body coherent photoemission process, including direct
        photoexcitation into a bulk state (1a), phonon-mediated photoexcitation
        into a bulk state (1b), and finally transmission of the excited bulk state into vacuum
        (2): incoming photon of momentum $\mathbf{K} \approx 0$ and energy
        $\hbar\Omega$ (dotted arrow), absorbed/emitted phonon of branch
        $\alpha$, momentum $\mathbf{k}_p$, and energy $\hbar\omega$ (jagged
        line), resulting electron in band $b'$ with momentum $\mathbf{k}$ or
        $\mathbf{k} \pm \mathbf{k}_p$ (filled circle), resulting hole in band
        $b$ with momentum $\mathbf{-k}$ (open circle), and 
        excited bulk-like state with vacuum components, consisting of 
        a Bloch wave in the far field of the material traveling toward the surface (solid arrow pointing away from the filled circle),
        a superposition of reflected Bloch waves in the far field of the material (downward solid arrows),
        and a superposition of plane waves in the far field of the vacuum (upward solid arrows), each of which has a momentum 
        $\mathbf{q} = \mathbf{k}_\parallel + \mathbf{G}_\parallel + \mathbf{q}_\perp$ 
        or, if a phonon is involved, $\mathbf{q} = (\mathbf{k} \pm \mathbf{k}_p)_\parallel + \mathbf{G}_\parallel 
        + \mathbf{q}_\perp$.
    }

    \label{fig:photoems}
\end{figure}

Figure~\ref{fig:photoems} illustrates the photoemission processes which we consider.
First, a photon excites an electron from an occupied bulk-like state which closely resembles a
bulk band $b$ and is bound to the material. The electron then transitions into an excited state that closely resembles a bulk band $b'$ in the material but, because the material is not infinite and has a surface, also has significant amplitude propagating in the vacuum. 
As Fig.~\ref{fig:photoems}(1) illustrates, this process may occur through either direct photoexcitation (1a), or phonon-mediated photoexcitation (1b) during which the
electron absorbs the photon while either coherently absorbing or emitting a phonon.
Our photoemission model thus resembles the one-step model\cite{ref:mahan, ref:ag-111_model} in that we consider coherent processes only, but goes beyond that model because we also include coherent electron-photon-phonon excitation processes.

The exact nature of the excited state plays an important role in our framework.  Far from the surface and deep into the material or far out into the vacuum, respectively, the excited state can be described as a superposition of pure Bloch waves or plane waves. As Fig.~\ref{fig:photoems}(2) illustrates, 
on the material side of the interface, the state will appear as a combination of the \textit{excited} Bloch state at bulk band $b'$
(which will will have a group velocity toward the surface if the electron is ultimately to be emitted) 
and a set of \textit{reflected} Bloch waves due to interaction with the surface. Similarly, far into the vacuum, the excited state will appear as a superposition of outgoing \textit{transmitted} plane waves.
The phases of all of the above superposed components must align at all points 
that are equivalent by the two-dimensional translational symmetry of the surface, 
and thus the crystal momentum component parallel to the 
surface must be conserved.\cite{ref:dowell} 
Specifically, for each plane-wave component $\ket{\mathbf{q}}$ of the outgoing wave in the far field of the vacuum, the parallel component of the wave vector $\mathbf{q}$
must match the sum of the parallel component of the excited Bloch state's crystal
momentum $\mathbf{k}_f$ 
and any reciprocal lattice vector $\mathbf{G}_s$ of the two-dimensional surface: $\mathbf{q}_{\parallel} =
\mathbf{k}_{f {\parallel}} + \mathbf{G}_s$. 
As Appendix~\ref{sec:app_Gpar} shows, for any three-dimensional bulk lattice and any surface, 
the surface reciprocal lattice vectors $\mathbf{G}_s$ correspond precisely to the projections $\mathbf{G}_\parallel$ onto the surface plane of all of the reciprocal lattice vectors $\mathbf{G}$ of the bulk crystal, allowing us to write
\begin{align}
    \mathbf{q}_\parallel = \mathbf{k}_{f \parallel} + \mathbf{G}_\parallel.
    \label{eqn:qpar}
\end{align}

In addition to the parallel component of the momentum, the total energy also must be
conserved during surface transmission,\cite{ref:dowell} so that the total kinetic energy $T(\mathbf{q})$
of the outgoing plane-wave component $\ket{\mathbf{q}}$ is
\begin{align}
    T(\mathbf{q}) = E_{\mathbf{k}_f,b'} - W,
    \label{eqn:Ktot}
\end{align}
\noindent where $E_{\mathbf{k}_f,b'}$ is the energy of the excited bulk state
$\ket{\mathbf{k}_f, b'}$ relative to the same reference with which the work function $W$ is determined.
Equations \eqref{eqn:qpar} and \eqref{eqn:Ktot} then yield the perpendicular component of the wave vector ${\mathbf{q}}$,
\begin{align}
    q_\perp (\mathbf{k}_f, b', \mathbf{G}_\parallel) 
        &\equiv \frac{\sqrt{2 m_e T_\perp^{\mathbf{k}_f, b', \mathbf{G}_\parallel} }}{\hbar},
    \label{eqn:qperp}
\end{align}
\noindent where the kinetic energy in the direction perpendicular to the surface is
\begin{align}
    T_\perp^{\mathbf{k}_f, b', \mathbf{G}_\parallel} 
        &= E_{\mathbf{k}_f,b'} - W - \frac{\hbar^2}{2m_e} |\kfPlusGPar|^2.
    \label{eqn:Kperp}
\end{align}

With the above ingredients in place, the mean-transverse energy is then the weighted average 
of the transverse kinetic energy of the plane-wave component $\ket{ \mathbf{q} (\mathbf{k}_f, b', \mathbf{G}_\parallel) }$ 
over all values of $\mathbf{k}_f$, $b'$, and $\mathbf{G}_\parallel$,
\begin{align}
    \text{MTE} (\Omega) = 
    \frac{
        \sum_{\mathbf{k}_f, b', \mathbf{G}_\parallel}
            \nu (\Omega, \mathbf{k}_f, b') 
                ~t (\mathbf{k}_f, b', \mathbf{G}_\parallel)
                ~T_{\parallel} (\mathbf{k}_{f \parallel}, \mathbf{G}_\parallel)
    }{
        \sum_{\mathbf{k}_f, b', \mathbf{G}_\parallel}
            \nu (\Omega, \mathbf{k}_f, b') 
                ~t (\mathbf{k}_f, b', \mathbf{G}_\parallel)
    },
    \label{eqn:MTE}
\end{align}
where the terms in the above expression are defined as follows. First, at photon energy $\hbar\Omega$ the photoexcitation transition rate $\nu(\Omega, \mathbf{k}_f, b')$
includes transitions from all possible initial bulk states to a particular excited bulk state $\ket{\mathbf{k}_f, b'}$ at momentum
$\mathbf{k}_f$ and band $b'$. 
Second, $t (\mathbf{k}_f, b', \mathbf{G}_\parallel)$ is the transmission probability 
for the excited bulk state $\ket{\mathbf{k}_f, b'}$ to ultimately emerge as the plane-wave $\ket{ \mathbf{q} (\mathbf{k}_f, b', \mathbf{G}_\parallel) }$ in vacuum.
Third, $T_\parallel (\mathbf{k}_{f \parallel}, \mathbf{G}_\parallel) 
= (\hbar^2 / 2m_e) | \kfPlusGPar |^2$ is the transverse kinetic energy of the plane-wave component
$\ket{ \mathbf{q}(\mathbf{k}_f, b', \mathbf{G}_\parallel) }$.
The next subsections describe the calculation of $\nu(\Omega, \mathbf{k}_f, b')$
and $t (\mathbf{k}_f, b', \mathbf{G}_\parallel)$, respectively.

Finally, we note that, from Eq.~\eqref{eqn:MTE}, it is apparent that
the transverse momentum distribution of the photoelectrons is the transverse momentum $\kfPlusGPar$ weighted by the product $\nu(\Omega, \mathbf{k}_f, b') ~t (\mathbf{k}_f, b', \mathbf{G}_\parallel)$.

%---------------------------------------------------------
%           Methods: Calculating Trans Rates
%---------------------------------------------------------
\subsection{Photoexcitation Transition Rates} \label{subsec:rates}

At a particular crystal momentum $\mathbf{k}$, 
a direct transition vertically excites an electron into a higher-energy
band at the same $\mathbf{k}$, because at photon energies of interest 
(few eVs),
the photon momenta are $\sim$10$^{3}$ smaller than the typical electronic
crystal momenta.
A phonon-mediated transition, however, excites an electron into a higher-energy
state at a different crystal momentum $\mathbf{k'} \neq \mathbf{k}$, because
unlike photons, phonons can have arbitrary crystal momenta.

Within the dipole approximation,~\cite{ref:arpes, ref:shankar_rates} 
first-order perturbation theory (Fermi's golden rule) and second-order perturbation theory give
the following transition rates of the direct photoexcitations and the phonon-mediated
photoexcitations,~\cite{ref:shankar_acs}
\begin{align}
    \nu_{\text{direct}}^{\mathbf{k},b'\leftarrow b} = \frac{2\pi}{\hbar} 
    \frac{e^2}{m_e^2} (f_{\mathbf{k},b}-f_{\mathbf{k},b'})
    \delta(E_{\mathbf{k},b'}-E_{\mathbf{k},b}-\hbar\Omega)
    \nonumber \\
    \times \left|
        (\mathbf{A} \cdot \mathbf{p})_{\mathbf{k},b' \leftarrow b}
    \right|^2
    \label{eqn:dir-exc_A}
\end{align}
\begin{align}
    \nu_{\text{phonon}}^{\mathbf{k'},b' \leftarrow \mathbf{k},b} 
    = &~\frac{2\pi}{\hbar}
    \frac{e^2}{m_e^2}
    (f_{\mathbf{k},b} - f_{\mathbf{k'},b'})
    \sum_{\alpha \pm} 
    \Bigg\{
        \left( n_{\mathbf{k'}-\mathbf{k},\alpha} + \frac{1}{2} \mp \frac{1}{2}
            \right)
    \nonumber \\
        &\times \delta(E_{\mathbf{k'},b'}-E_{\mathbf{k},b}-\hbar\Omega \mp
            \hbar\omega_{\mathbf{k'}-\mathbf{k},\alpha})
    \nonumber \\
        &\times \left| \sum_{m} \left(
            \frac{g_{\mathbf{k'},b' \leftarrow \mathbf{k},m}^{\mathbf{k'}-\mathbf{k},\alpha}
                  \left( \mathbf{A} \cdot \mathbf{p} \right)_{\mathbf{k},m \leftarrow b} 
            }
            {E_{\mathbf{k},m}-E_{\mathbf{k},b}-\hbar\Omega + i\eta_{\mathbf{k},m}
            } \right. \right.
    \nonumber \\
    &\left. \left.
       +\frac{\left( \mathbf{A} \cdot \mathbf{p} \right)_{\mathbf{k'},b' \leftarrow m}
              g_{\mathbf{k'},m \leftarrow \mathbf{k},b}^{\mathbf{k'}-\mathbf{k},\alpha}
        }
        {E_{\mathbf{k'},m}-E_{\mathbf{k},b} \mp 
            \hbar\omega_{\mathbf{k'}-\mathbf{k},\alpha} + i\eta_{\mathbf{k'},m}
        }
    \right) \right|^2 \Bigg\},
    \label{eqn:phn-exc_A}
\end{align}
where the relevant quantities are defined as follows. 
The indices $b$, $b'$, and $\alpha$ label the initial bulk band, excited bulk band, and phonon branch, respectively, and the constants $e$ and $m_e$ are the electron charge and vacuum electron mass.
The quantities $f$, $E$, $n$, $\omega$, and $\Omega$ respectively are the
electron Fermi occupancy, electron band energy, phonon Bose occupancy, phonon
frequency, and photon frequency, and the $\mp$ sign labels phonon absorption/emission respectively.
The matrix element $(\mathbf{A} \cdot \mathbf{p})_{\mathbf{k}, j \leftarrow i} \equiv
\braket{\mathbf{k},j | \mathbf{A} \cdot \mathbf{p} | \mathbf{k},i}$ is the
electron-photon interaction matrix element, 
and $g_{\mathbf{k'},j \leftarrow \mathbf{k},i}^{\mathbf{k'}-\mathbf{k},\alpha}$
is the electron-phonon interaction matrix element between an initial electronic state
$\ket{\mathbf{k},i}$, a phonon of momentum $\mathbf{k'}-\mathbf{k}$ and branch
$\alpha$, and a final electronic state $\ket{\mathbf{k'},j}$. 
Finally, 
$\eta_{\mathbf{k},m}$ is the electron linewidth of the intermediate state of the
phonon-mediated photoexcitation at a particular momentum $\mathbf{k}$ and band $m$.

The electron linewidth $\eta_{\mathbf{k},m}$ considered in this work 
is the sum of the contributions to the imaginary part of the self-energy from electron-electron and electron-phonon scattering
at the intermediate state $\ket{\mathbf{k},m}$,
\begin{align}
    \eta_{\mathbf{k},m} = \operatorname{Im}\Sigma_{\mathbf{k},m}^\text{e--e} 
    + \operatorname{Im}\Sigma_{\mathbf{k},m}^\text{e--ph},
    \label{eqn:eta}
\end{align}
where Ref.~\onlinecite{ref:shankar_acs} gives the expressions for $\operatorname{Im}\Sigma_{\mathbf{k},m}^\text{e--e}$ and
$\operatorname{Im}\Sigma_{\mathbf{k},m}^\text{e--ph}$.

Choosing the Coulomb gauge ($\nabla \cdot \mathbf{A} = 0$)~\cite{ref:GnY} to
quantize the vector potential $\mathbf{A}$ reduces
Equations~\eqref{eqn:dir-exc_A} and~\eqref{eqn:phn-exc_A} to
\begin{align}
    \nu_{\text{direct}}^{\mathbf{k},b'\leftarrow b} = 
    ~&\frac{2\pi}{\hbar} \frac{e^2}{m_e^2} |A_0 (\Omega)|^2
    (f_{\mathbf{k},b}-f_{\mathbf{k},b'})
    \delta(E_{\mathbf{k},b'}-E_{\mathbf{k},b}-\hbar\Omega)
    \nonumber \\
    &\times \left|
        \hat{\boldsymbol{\upepsilon}}(\Omega) \cdot \mathbf{p}_{\mathbf{k},b' \leftarrow b}
    \right|^2
    \label{eqn:dir-exc}
\end{align}
and
\begin{align}
    \nu_{\text{phonon}}^{\mathbf{k'},b' \leftarrow \mathbf{k},b} = 
    ~&\frac{2\pi}{\hbar} \frac{e^2}{m_e^2} |A_0 (\Omega)|^2
    (f_{\mathbf{k},b} - f_{\mathbf{k'},b'})
    \nonumber \\
    &\times \sum_{\alpha \pm} 
    \Bigg\{
        \left( n_{\mathbf{k'}-\mathbf{k},\alpha} + \frac{1}{2} \mp \frac{1}{2}
            \right)
        \nonumber \\
        &\times \delta(E_{\mathbf{k'},b'}-E_{\mathbf{k},b}-\hbar\Omega \mp
            \hbar\omega_{\mathbf{k'}-\mathbf{k},\alpha})
        \nonumber \\
        &\times \left| \hat{\boldsymbol{\upepsilon}}(\Omega) \cdot \sum_{m} \left(
            \frac{g_{\mathbf{k'},b' \leftarrow \mathbf{k},m}^{\mathbf{k'}-\mathbf{k},\alpha}
                  ~\mathbf{p}_{\mathbf{k},m \leftarrow b}
            }
            {E_{\mathbf{k},m}-E_{\mathbf{k},b}-\hbar\Omega + i\eta_{\mathbf{k},m}
            } \right. \right.
        \nonumber \\
        &\left. \left.
            +\frac{ \mathbf{p}_{\mathbf{k'},b' \leftarrow m}
                  ~g_{\mathbf{k'},m \leftarrow \mathbf{k},b}^{\mathbf{k'}-\mathbf{k},\alpha}
            }
            {E_{\mathbf{k'},m}-E_{\mathbf{k},b} \mp 
                \hbar\omega_{\mathbf{k'}-\mathbf{k},\alpha} + i\eta_{\mathbf{k'},m}
            }
        \right) \right|^2
    \Bigg\},
    \label{eqn:phn-exc}
\end{align}
where $\mathbf{p}_{\mathbf{k}, j \leftarrow i} \equiv \braket{\mathbf{k},j | \mathbf{p} | \mathbf{k},i}$ is the momentum operator matrix element and
$\hat{\boldsymbol{\upepsilon}}(\Omega)$ is the polarization unit vector of the photon with energy $\hbar\Omega$
inside the bulk of the material. Finally, $A_0(\Omega)$ is the amplitude of the vector potential $\mathbf{A}$ at photon energy $\hbar\Omega$, which does not contribute to the final distribution of photoelectrons because of the normalization factor in Eq.~\eqref{eqn:MTE}.

At a particular photon energy $\hbar\Omega$, depending on the polarization of the incident laser photons in vacuum, the photons inside the material can have multiple polarizations that sum incoherently. 
Therefore, in general, the
total rate $\nu (\Omega, \mathbf{k}_f, b')$ of all transitions into a bulk
excited state $\ket{\mathbf{k}_f, b'}$ involving photons with energy
$\hbar\Omega$ is 
\begin{align}
    \nu (\Omega, \mathbf{k}_f, b') 
    = \sum_{\hat{\boldsymbol{\upepsilon}}(\Omega)} 
       a( \hat{\boldsymbol{\upepsilon}}(\Omega) ) 
       ~\Big\{
           &\nu_d (\Omega, \mathbf{k}_f, b'; \hat{\boldsymbol{\upepsilon}(\Omega)}) 
           \nonumber
           \\
           &+ \nu_p (\Omega, \mathbf{k}_f, b'; \hat{\boldsymbol{\upepsilon}(\Omega)})
       \Big\},
    \label{eqn:rates_fin}
\end{align}
where $a( \hat{\boldsymbol{\upepsilon}}(\Omega) )$ is the weight of the photons with an energy $\hbar\Omega$ and a particular polarization $\hat{\boldsymbol{\upepsilon}}(\Omega)$ inside the material,
$\nu_d (\Omega, \mathbf{k}_f, b'; \hat{\boldsymbol{\upepsilon}}(\Omega)) \equiv \sum_b
\nu_{\text{direct}}^{\mathbf{k}_f, b' \leftarrow b}$ 
is the direct transition rate Eq.~\eqref{eqn:dir-exc} summed over all initial bands, and 
$\nu_p (\Omega, \mathbf{k}_f, b'; \hat{\boldsymbol{\upepsilon}}(\Omega)) \equiv
\sum_{\mathbf{k} b} \nu_{\text{phonon}}^{\mathbf{k}_f, b' \leftarrow
\mathbf{k},b}$
is the phonon-mediated transition rate Eq.~\eqref{eqn:phn-exc} summed over all initial states.

Our experiments on PbTe(111) use \textit{unpolarized} laser light  
with an angle of incidence $\approx$ 35$^\circ$
and with an undetermined azimuthal direction with respect to the underlying crystalline axes. 
To deal with the latter uncertainty, we have considered both of 
the two distinct high-symmetry incoming laser beam directions along the surface, $\pm [ 1 \bar{1} 0 ]$ and $\pm [ 1 1 \bar{2} ]$, ultimately finding very similar results.
For any given incoming beam direction there is a unique $s$-polarization direction and a $p$-polarization direction inside the material determined by the angle of refraction, which 
at the photon energies of interest ($\sim$4--5~eV) vary in our material between 40$^\circ$ and 70$^\circ$.
We find that for both of our considered incoming directions, the corresponding $s$-polarization and the three $p$-polarizations we have considered (corresponding to the refracted angles of 40$^\circ$, 60$^\circ$, and 70$^\circ$) all yield quantitatively similar results for the MTE as a function of photon energy at the photon energies of interest. Accordingly, for this work, we choose to focus on a single, representative laser light in the material consisting of an equal, incoherent mixture of
an $s$-polarization in the $\pm [ 1 \bar{1} 0 ]$ direction and 
a $p$-polarization corresponding to a refracted angle of 60$^\circ$.

%---------------------------------------------------------
%          Methods: Calculating Surf Trans Prob
%---------------------------------------------------------
\subsection{Surface Transmission Probability} \label{subsec:surfTrans}

Two conditions determine whether an electron in an excited bulk state transmits into vacuum.
First, the electron's group velocity must be in the direction toward the surface.
Second, to avoid total internal reflection, the electron must couple to 
the plane-wave components in the far field of the vacuum
that have positive kinetic energies in the direction perpendicular to the surface.
As discussed in Sec.~\ref{subsec:framework}, surface transmission conserves
both the total energy and the momentum component parallel to the surface,~\cite{ref:dowell} 
and thus it is possible for the kinetic energy perpendicular to the surface to become negative. 
With these considerations and including the appropriate kinematic factors, the transmission probability $t (\mathbf{k}_f, b', \mathbf{G}_\parallel)$ from the bulk state $\ket{\mathbf{k}_f, b'}$ to the vacuum plane-wave component $\ket{ \mathbf{q} (\mathbf{k}_f, b', \mathbf{G}_\parallel) }$ is then
\begin{align}
    t (\mathbf{k}_f, b', \mathbf{G}_\parallel) 
        =
            &~\Theta \left( \mathbf{v}_\text{group}^{\mathbf{k}_f,b'} 
                                  \cdot \hat{\mathbf{n}} 
                           \right)
            ~\Theta \left( T_\perp^{\mathbf{k}_f, b', \mathbf{G}_\parallel} \right)
            \nonumber
            \\
            &\times | D_{\mathbf{k}_f, b', \mathbf{G}_\parallel} |^2
            \frac{ \left[ 
                         (2/m_e) ~T_\perp^{\mathbf{k}_f, b', \mathbf{G}_\parallel} 
                   \right]^{1/2} }
                 { \mathbf{v}_\text{group}^{\mathbf{k}_f,b'} \cdot \hat{\mathbf{n}} },
    \label{eqn:trans-prob-per-G_exact}
\end{align}
where $\mathbf{k}_f$ and $b'$ respectively label the crystal momentum
and the band of the excited bulk state, and
$\mathbf{G}_\parallel$ is the surface projection of the vector $\mathbf{G}$ in the reciprocal bulk lattice.
In the above equation,
$\Theta(x)$ is the Heaviside step function, 
$D_{\mathbf{k}_f, b', \mathbf{G}_\parallel}$ is the amplitude of the plane-wave
component $\ket{\mathbf{q}(\mathbf{k}_f, b', \mathbf{G}_\parallel)}$ in vacuum,
$\mathbf{v}_\text{group}^{\mathbf{k}_f,b'} \equiv (1/\hbar)
\mathbf{\nabla}_{\mathbf{k}_f} E_{\mathbf{k}_f,b'}$ is the group velocity of
the excited electron in the material, 
$\hat{\mathbf{n}}$ is the unit vector of the surface normal (which is parallel
to $[111]$ in this work), 
and $T_\perp^{\mathbf{k}_f, b', \mathbf{G}_\parallel}$ is given by Eq.~\eqref{eqn:Kperp}.
The fraction in the above equation gives the appropriate kinematic factors as the ratio of the perpendicular group velocity of the plane-wave component $\ket{\mathbf{q}(\mathbf{k}_f, b', \mathbf{G}_\parallel)}$ in vacuum to the perpendicular group velocity of the excited bulk state $\ket{\mathbf{k}_f, b'}$ in the material. 

The proper way of calculating $D_{\mathbf{k}_f, b', \mathbf{G}_\parallel}$, 
the amplitude of the plane-wave component
$\ket{\mathbf{q}(\mathbf{k}_f, b', \mathbf{G}_\parallel)}$ in the far field of the vacuum, 
is to employ an \abinit scattering theory to relate 
$D_{\mathbf{k}_f, b', \mathbf{G}_\parallel}$
with the amplitudes of the plane-wave components of the excited Bloch state 
$\ket{\mathbf{k}_f, b'}$ in the far field of the material. 
Such a theory requires solving the three-dimensional Schr\"{o}dinger's equation \textit{from first principles}, with 
techniques that result in eigenstates corresponding to a potential from a semi-infinite slab of material and a semi-infinite slab of vacuum.

For simplicity, we have considered approximating $D_{\mathbf{k}_f, b',
\mathbf{G}_\parallel}$ in three different ways,
\begin{subnumcases}{| D_{\mathbf{k}_f, b', \mathbf{G}_\parallel} |^2 \approx }
    &\hspace{-1.5em}$| \sum_{\mathbf{G}_\perp} C_{\mathbf{k}_f, b', 
                                                  \mathbf{G}_\parallel + \mathbf{G}_\perp} |^2$
    \label{eqn:D-approxes_cohSum}
    \\
    &\hspace{-1.5em}$\max \{ | C_{\mathbf{k}_f, b', 
                              \mathbf{G}_\parallel + \mathbf{G}_\perp} |^2
                          \}_{\mathbf{G}_\perp}$
    \label{eqn:D-approxes_max}
    \\
    &\hspace{-1.5em}$|D|^2$,
    \label{eqn:D-approxes_const}
\end{subnumcases}
where $D$ is a non-zero constant and $C_{\mathbf{k}_f, b',
\mathbf{G}}$ is the amplitude of the plane-wave component of
the excited bulk Bloch state associated with the reciprocal lattice vector $\mathbf{G}$, so that the bulk Bloch state $\ket{\mathbf{k}_f, b'}$ in \textit{real space} is 
$\psi_{\mathbf{k}_f, b'}(\mathbf{r}) =
\sum_\mathbf{G} C_{\mathbf{k}_f, b', \mathbf{G}} \exp{[i (\mathbf{k}_f +
\mathbf{G}) \cdot \mathbf{r}]}$. The motivations for these approximations are as follows.
Equation~\eqref{eqn:D-approxes_cohSum} assumes that the plane-wave
component $\ket{\mathbf{q} (\mathbf{k}_f, b', \mathbf{G}_\parallel)}$ 
equals the superposition of the plane-wave components of the
excited Bloch state $\ket{\mathbf{k}_f, b'}$ that have the same $\mathbf{G}_\parallel$.
Equation~\eqref{eqn:D-approxes_max} assumes that for each 
$\mathbf{G}_\parallel$, only the plane-wave component of $\ket{\mathbf{k}_f, b'}$
with the largest probability determines $\ket{\mathbf{q} (\mathbf{k}_f, b', \mathbf{G}_\parallel)}$.
Lastly, Equation~\eqref{eqn:D-approxes_const} is the simple constant-amplitude approximation.

For PbTe(111), all three approximations give quantitatively similar results for the MTE as a function of photon energy at photon energies of interest 
($\sim$4--5~eV).
Accordingly, the results presented in this work use
Eq.~\eqref{eqn:D-approxes_const}, the simplest approximation of the three, so that from here forward we use
\begin{align}
    t (\mathbf{k}_f, b', \mathbf{G}_\parallel) 
        \approx 
            &~\Theta \left( \mathbf{v}_\text{group}^{\mathbf{k}_f,b'} 
                                  \cdot \hat{\mathbf{n}} 
                           \right)
            ~\Theta \left( T_\perp^{\mathbf{k}_f, b', \mathbf{G}_\parallel} \right)
            \nonumber
            \\
            &\times | D |^2
            \frac{ \left[ 
                         (2/m_e) ~T_\perp^{\mathbf{k}_f, b', \mathbf{G}_\parallel} 
                   \right]^{1/2} }
                 { \mathbf{v}_\text{group}^{\mathbf{k}_f,b'} \cdot \hat{\mathbf{n}} }.
    \label{eqn:trans-prob-per-G}
\end{align}

%---------------------------------------------------------
%           Methods: Computational Details
%---------------------------------------------------------
\subsection{Computational Details} \label{subsec:details}

Calculation of MTE (Eq.~\eqref{eqn:MTE})
involves evaluating high-dimensional sums over continuous set of crystal momenta $\{ \mathbf{k} \}$ using the Monte Carlo method.
Here, we employ the the Wannier interpolation method\cite{ref:wannier_review} to
efficiently interpolate the required quantities, such as electron linewidths and matrix elements, for arbitrary values of $\mathbf{k}$.
This interpolation method involves expressing such quantities in a maximally-localized Wannier basis formed by linear combinations of Bloch wavefunctions,\cite{ref:wannier, ref:wannier_review} which requires calculation of the electronic structure of the material.

To calculate the electronic structure of PbTe, we employ the plane-wave density-functional theory (DFT) framework, 
with the GGA-PBE exchange correlation functional\cite{ref:pbe} and fully-relativistic
norm-conserving pseudopotentials\cite{ref:oncv} from the SG15 library,\cite{ref:sg15} as implemented in the JDFTx software framework.\cite{ref:jdftx}  
We have found that using non-relativistic and scalar-relativistic pseudopotentials for PbTe yields a band gap that exceeds the experimental gap. Solving this problem requires spin-orbit coupling,\cite{ref:pbte_soc} which necessitates fully-relativistic pseudopotentials. 

The calculations of the bulk electronic structure employ a face-centered-cubic (FCC) primitive cell of PbTe, a plane-wave cutoff of 20 Hartrees, a Brillouin zone sampling mesh of $6 \times 6 \times 6$, and an optimized PbTe lattice constant of 6.57 \AA~(within 2\% from the experimental value\cite{ref:pbte_exp-latt-const}). To deal with the classic DFT band-gap problem, we scissor the conduction band energies to match the experimental gap of 0.3 eV.\cite{ref:pbte_exp-gap} 
Using linear combinations of the bulk Bloch bands at energies from $-4.8$~eV below to  12.7~eV above the valence band maximum, we generate a maximally-localized Wannier basis set using a supercell of $6 \times 6 \times 6$ FCC primitive cells. 
This Wannier basis set reproduces the bulk band structure
at the energy range from $-4.8$~eV below to 5.7~eV above the valence band maximum, which is sufficient to include all photoexcitations with photon energies of interest ($\sim$4--5~eV).
The determination of these Wannier functions and their use below in determining linewidths, matrix elements and MTEs are all based on the implementation of JDFTx\cite{ref:jdftx} described in Ref.~\onlinecite{ref:shankar_acs}.

Because this work considers phonon-mediated photoexcitations, we also calculate the force matrix for bulk phonons and the electron-phonon matrix elements of PbTe using a modified version of the frozen phonon method, as implemented in
JDFTx,\cite{ref:jdftx} which
allows calculations of phonons at arbitrary wave-vectors. These calculations use a supercell of $3\times 3\times 3$ FCC primitive 
cells and DFT parameters corresponding to those of the bulk electronic structure calculations described above.

The phonon-mediated excitations also require calculation of the electron linewidths (Eq.~\eqref{eqn:eta}) for all bulk electronic states.
The methods of these calculations are detailed elsewhere.\cite{ref:shankar_acs}
For this work, the calculation of the electron-electron scattering contribution to the linewidth uses a frequency grid resolution of 0.001~eV and a cutoff of $\sim$130~eV for the dielectric matrices.
The calculation of the electron-phonon scattering contribution uses a fine wave-vector grid of $168 \times 168 \times 168$.

The surface transmission probability of a photoexcited electron (Eq.~\eqref{eqn:trans-prob-per-G}) depends on the work function of the material surface.
Although it is possible to calculate the work function of PbTe(111) \textit{ab initio},\cite{ref:li-sch-p3, ref:li-sch} the effective work function in our experiments can 
be quite different due to effects such as the surface condition of our sample
and the Schottky effect.\cite{ref:li-sch, ref:alk-ant_cryo} Because careful measurements of the effective work function of our sample are not
available, we determine the effective work function by comparing our calculated
MTEs with our experimental MTEs, finding that a work function of 4.05 eV (at the low end of the range 4.1--4.9~eV reported in the literature~\cite{ref:li-sch-p3, ref:li-sch, ref:pbte_workFunc_spicer, ref:pbte_workFunc_weiser, ref:pbte_workFunc_basu}) results in the best agreement between 
our calculated and experimental MTEs
as functions of photon energy.

Finally, to determine the MTEs, this work uses $\sim$1$.5 \times 10^8$ Monte Carlo samples of crystal momenta $\mathbf{k}$ to converge our results in the photon energy range of interest ($\sim$4--5~eV).
To enumerate the possible outgoing plane waves, we include the $\mathbf{G}_\parallel$ vectors for the nearest neighbors of $\mathbf{G}_\parallel = 0$, the origin of the surface-projected reciprocal lattice. This range of $\mathbf{G}_\parallel$ is sufficient to cover the transverse momenta measured in our experiments.

%===============================================================
%                   END - METHODS
%===============================================================

\section{Results and Discussion} \label{sec:results}

%---------------------------------------------------------
%                   Results (A): MTE
%---------------------------------------------------------
\subsection{Mean Transverse Energies} \label{subsec:MTEresult}

\begin{figure}[h!]
    \centering
    \includegraphics[width=0.9\linewidth]{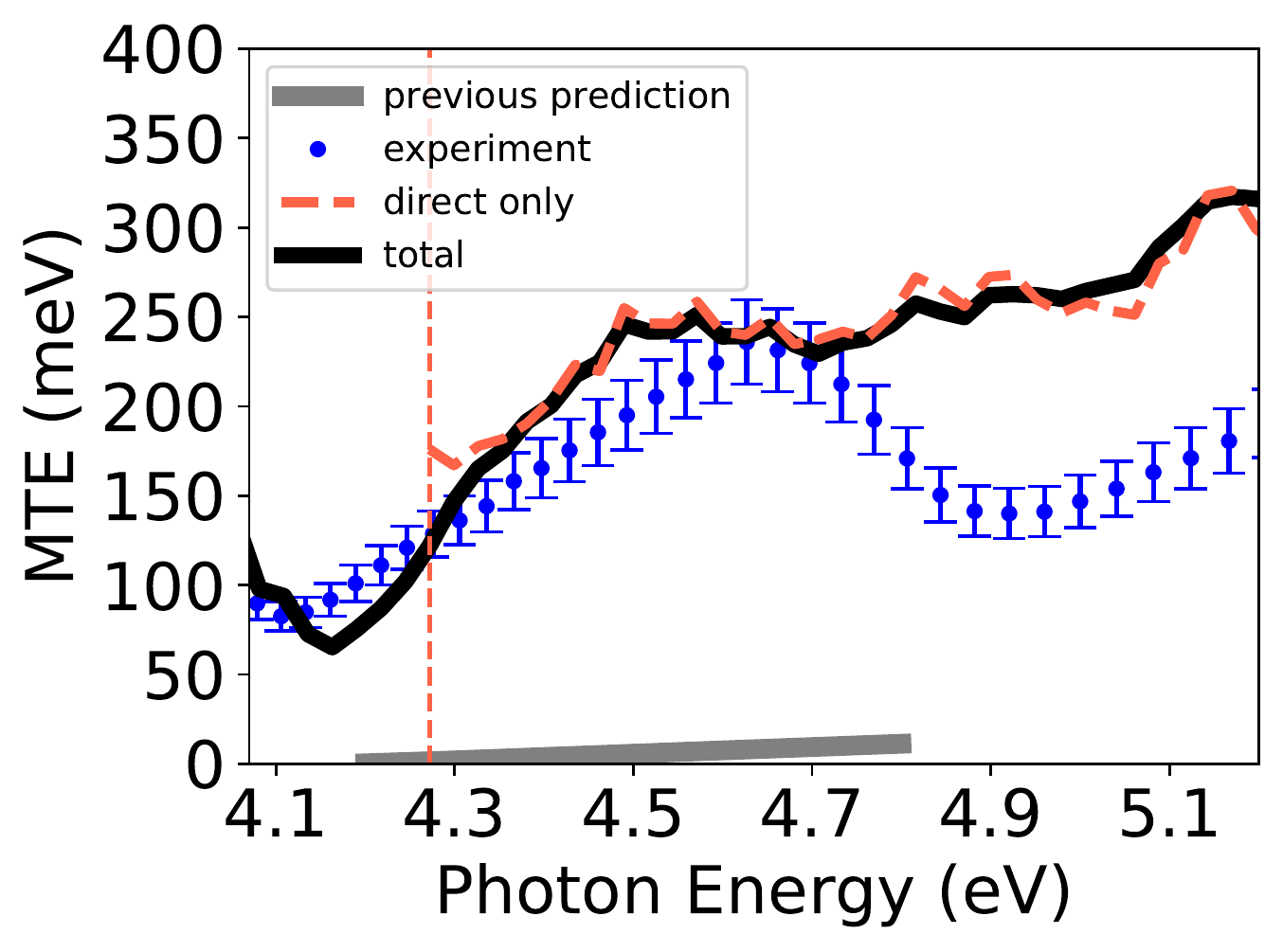}

    \caption{
        Mean transverse energy (MTE) of photoelectrons emitted from PbTe(111) as a function of laser photon energy at room temperature: 
        our experimental results (points with error bars),
        our calculations including direct and phonon-mediated excitations (solid curve),
        our calculations including solely direct excitations (dashed curve),
        previous predictions~\cite{ref:li-sch, ref:li-sch-p3} (thick curve with small MTE values),
        and our calculated threshold of direct photoexcitation (vertical dashed line).
        Compared to previous predictions, which are an order of magnitude smaller, our calculations yield far better agreement with our experiments.
    }

    \label{fig:MTE-298K}
\end{figure}

Figure~\ref{fig:MTE-298K} shows our results for the mean transverse energy of emitted photoelectrons as a function of laser photon energy and compares them with previous predictions\cite{ref:li-sch-p3, ref:li-sch} and our experimental measurements.
First, we find that including both the direct and phonon-mediated photoexcitations into bulk-like states, as opposed to vacuum states, reproduces the magnitude and general trends of the measured MTEs both below and above the calculated direct threshold of 4.27~eV. 
Although we do not reproduce the experimentally observed dip centered at 4.9~eV, we do find a leveling off of the increase in the MTE at similar energies. Section~\ref{subsec:momentotron}
explores this discrepancy in more detail, showing that this feature is likely due to photoexcitations directly into vacuum states.
Second, we find that including only the direct excitations reproduces relatively well the MTEs above the calculated threshold, although the \textit{indirect} excitations above threshold actually account for at least $\sim$45\% of the total number of emitted photoelectrons. 
Both the need to consider phonon-mediated processes below the direct threshold {\em and} the prevalence of such processes above threshold underscore the importance of indirect transitions due to phonon effects in photoemission from PbTe(111).

\begin{figure}[h!]
    \centering
    \includegraphics[width=0.9\linewidth]{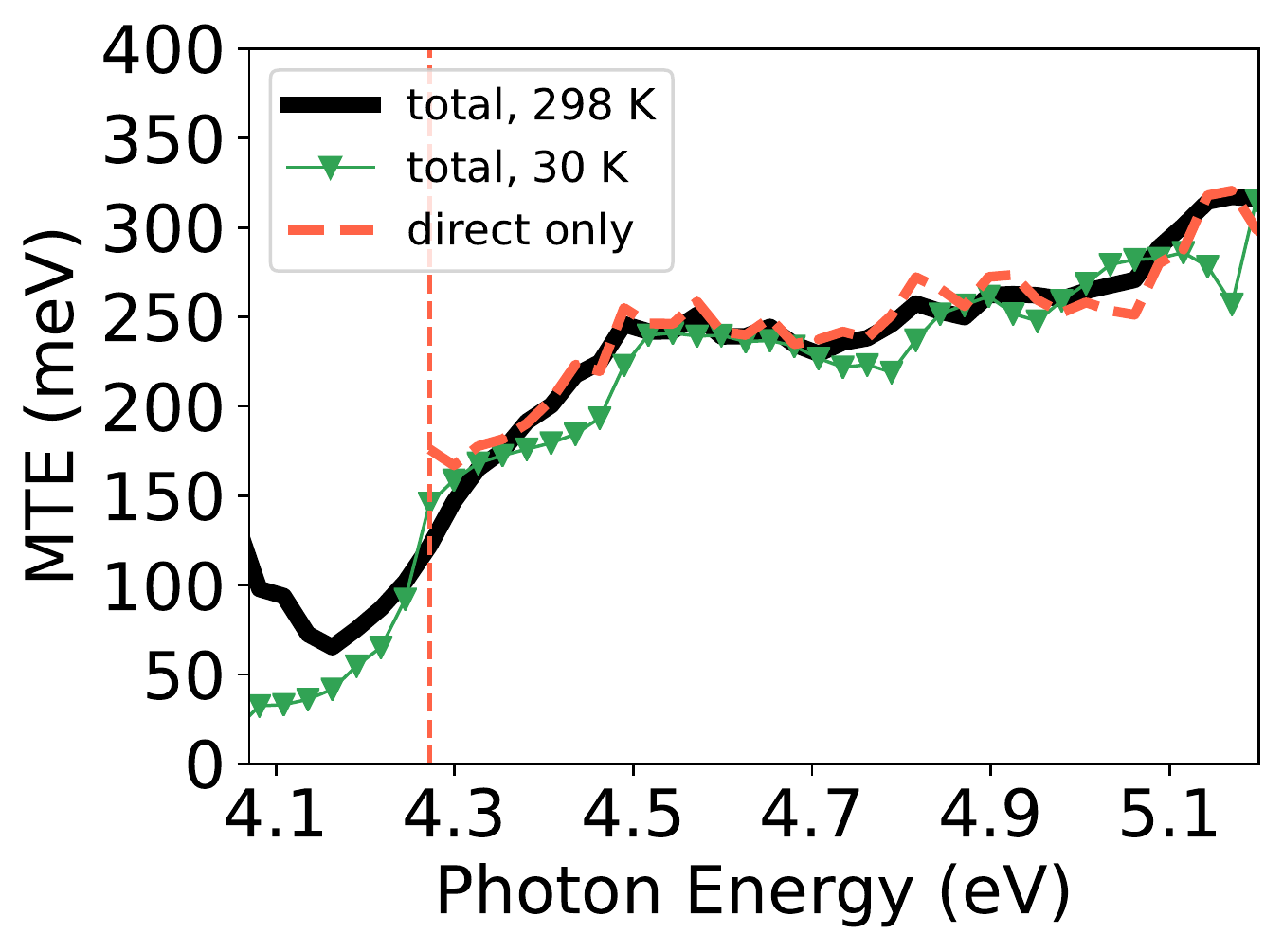}

    \caption{
        Calculated MTE as a function of laser photon energy: 
        with phonon effects at room temperature (solid curve), 
        with phonon effects at 30~K (downward triangles), 
        without phonon effects (dashed curve), 
        and the threshold of direct photoexcitation (vertical dashed line).
    }
    \label{fig:MTE-298K-30K-dir}
\end{figure}

The significance of the above phonon effects suggests that 
operation at cryogenic temperatures might reduce the MTEs.
To explore such effects, Fig.~\ref{fig:MTE-298K-30K-dir} 
contrasts our room-temperature results with what we predict for the MTEs at 30~K. 
Above the direct threshold of 4.27~eV, the MTEs at 30~K are approximately equal
to the MTEs at room temperature as well as the MTEs due to
direct processes only, which are not affected significantly by temperature.
Below the direct threshold, we indeed predict a lowering of the MTEs when operating at 30~K. However, unlike in polycrystalline metallic photocathodes where the MTE 
is directly proportional to the thermal energy $k_B T$ near and below threshold,\cite{ref:mte_thermal-limit} for single-crystal PbTe(111) 
we find a more complicated behavior and not nearly the expected factor of ten reduction. Simply lowering the operating temperature of single-crystal photocathodes is not guaranteed to provide a significant reduction in the MTE.

Beyond not necessarily providing significant reduction in MTE, lowering the photocathode temperature reduces the number of phonons available for indirect photoexciatation processes and
thus may lower the quantum efficiency, thereby actually 
reducing the overall beam brightness. Without lowering of the photocathode temperature as a guaranteed method, improving beam brightness from single-crystal photocathodes will require materials whose band structures allow photoexcitations of electrons with low transverse momenta even at room temperature. Discovery of such materials requires a deeper understanding of the transverse momentum distributions of emitted photoelectrons, which the next section explores.

%---------------------------------------------------------
%               Results (B): Trans Mom Distrib
%---------------------------------------------------------
\subsection{Transverse Momentum Distributions} \label{subsec:momentotron}

\begin{figure*}[t!]
    \centering
    \includegraphics[width=\linewidth]{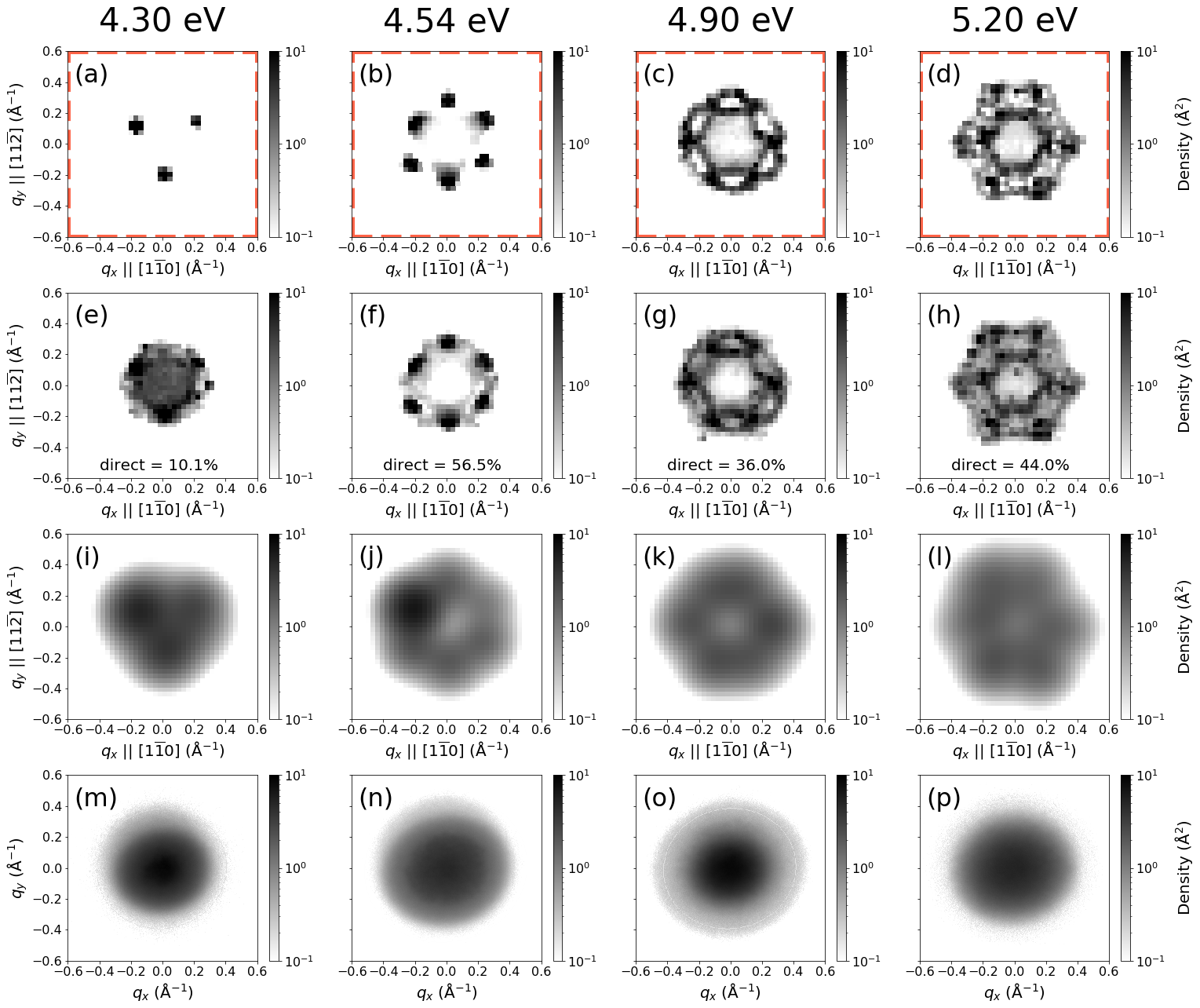}

    \caption{ 
        Transverse momentum distributions, normalized as probability densities,
        on the $q_x q_y$ plane parallel to the (111) surface at four different
        photon energies near and above the direct threshold:
        distributions including only direct excitations (top row), 
        distributions including both direct and phonon-mediated excitations (second row), 
        fractions, in percents, of the probability densities that are due to direct excitations (second row, ``direct''), 
        distributions including both direct and phonon-mediated excitations with Gaussian smearing of RMS width 0.1 \AA$^{-1}$ (third row), 
        and experimentally measured distributions (bottom row).
        The overall sizes and the trends of the distributions are consistent with the MTE data in Fig.~\ref{fig:MTE-298K}.
    }

    \label{fig:trans-mom-distrib}
\end{figure*}

To further elucidate the photoemission process and to explore the origin of the lack of the dip near 4.9~eV in our predicted mean transverse energies (MTEs), we now explore the detailed transverse momentum distribution of the photoemitted electrons.
This distribution is a two-dimensional histogram of the transverse momenta
$\kfPlusGPar$ of all emitted photoelectrons in the plane parallel to the surface.  As discussed in Sec.~\ref{subsec:framework}, we can extract this histogram from our framework by taking the histogram weights to equal
the product of the photoexcitation transition rate $\nu (\Omega, \mathbf{k}_f, b')$
(Eq.~\eqref{eqn:rates_fin}) and the surface transmission probability $t
(\mathbf{k}_f, b', \mathbf{G}_\parallel)$ (Eq.~\eqref{eqn:trans-prob-per-G}).
Figure~\ref{fig:trans-mom-distrib} shows the room-temperature transverse momentum distributions at photon energies near and above the direct threshold of 
4.27~eV, comparing the results from our calculations under various approximations with our experimental results. 

The first two rows of Fig.~\ref{fig:trans-mom-distrib} compare the results from considering the direct processes only with the results from including also the indirect processes. As indicated in the second row, although the indirect processes make a significant contribution, neither the indirect nor the direct processes completely dominate.
Furthermore, we see that at the photon energies considered, above 4.3~eV
the transverse momentum distributions from the direct-only processes are
similar to the distributions from the combined direct and indirect processes,
explaining why both the direct-only MTEs and the total MTEs are
approximately equal above the direct threshold (Fig.~\ref{fig:MTE-298K-30K-dir}). 
Finally, it is apparent from Fig.~\ref{fig:trans-mom-distrib} that both the direct and phonon-mediated photoexcitations result in photoelectrons with primarily significant transverse momenta, thereby corresponding to the large calculated MTEs of a few hundred meV that we find above the direct threshold.

%\begin{figure}[h!]
%\end{figure}

Figure~\ref{fig:trans-mom-distrib} also compares our calculated distributions with our experimental results shown on the bottom-most row. 
Note that, unlike the calculated distributions on the second row, the experimental distributions \textit{do not} show the three-fold 
symmetry of the (111) surface of PbTe, but show nearly cylindrical symmetry. 
We believe this is \textit{not} due to polycrystallinity, because our sample shows a clear
hexagonal pattern from low-energy electron diffraction experiments. 
The cylindrical smearing is then likely due to other effects such as non-uniform electric fields on the photocathode surface. 
Possible causes of these fields include surface relaxations and reconstructions, as well as small rough patches and atomic steps, all of which have been experimentally observed on PbTe(111).\cite{ref:pbte111_recons_exp, ref:wjid-hines_pbte111-stm}

To account for the observed cylindrical smearing effects in a simplified way, the third row of 
Fig.~\ref{fig:trans-mom-distrib} shows the results from the second row convolved with a two-dimensional Gaussian of RMS width 0.1
\AA$^{-1}$, which gives the best overall agreement 
with the experimental distributions.
The convolved distributions have similar 
sizes to the experimental distributions up until 5.2~eV, where the convolved distribution is noticeably larger than that observed experimentally, consistent with the larger predicted MTE in Fig.~\ref{fig:MTE-298K}.

Contributing to our overprediction of MTEs at high photon energies is the fact that our convolved transverse momentum distributions remain somewhat ``hollow'' with low contributions in the center, as contrasted with our measured distributions which tend to be peaked at the center.
One possible explanation for this difference
is that our calculations exclude contributions from the
photoelectrons that transition directly into vacuum states.
As explained in Sec.~\ref{sec:prev-vs-exp}, 
among such electrons there are significant contributions from transitions along
the $\Gamma$--L direction with zero transverse momenta, which would tend to fill in the distributions and lower the predicted MTE. We believe that future work combining bulk-like transitions with transitions directly into vacuum will further improve the agreement between theory and experiment, and, in particular, will reproduce the dip in MTE near 4.9~eV observed in Sec.~\ref{subsec:MTEresult}.

%===============================================================
%                   END - RESULT
%===============================================================

\section{Computational Search for low-MTE single-crystal photocathodes} \label{sec:search}

The insights gained in the previous sections enable the development of an efficient computational screening procedure to search for single-crystal materials that yield photoelectrons with low mean transverse energies.
Such screening must consider both excitation into bulk-like states and also excitation directly into vacuum, both of which must yield low MTEs.
References \onlinecite{ref:li-sch-p3} and \onlinecite{ref:li-sch} give an example of 
screening based on direct excitations into vacuum. 
The remainder of this section focuses on important considerations when screening based on excitations into bulk-like states. 

\begin{figure*}[t!]
%\begin{figure}[h!]
    \centering

    \includegraphics[height=0.4\linewidth]{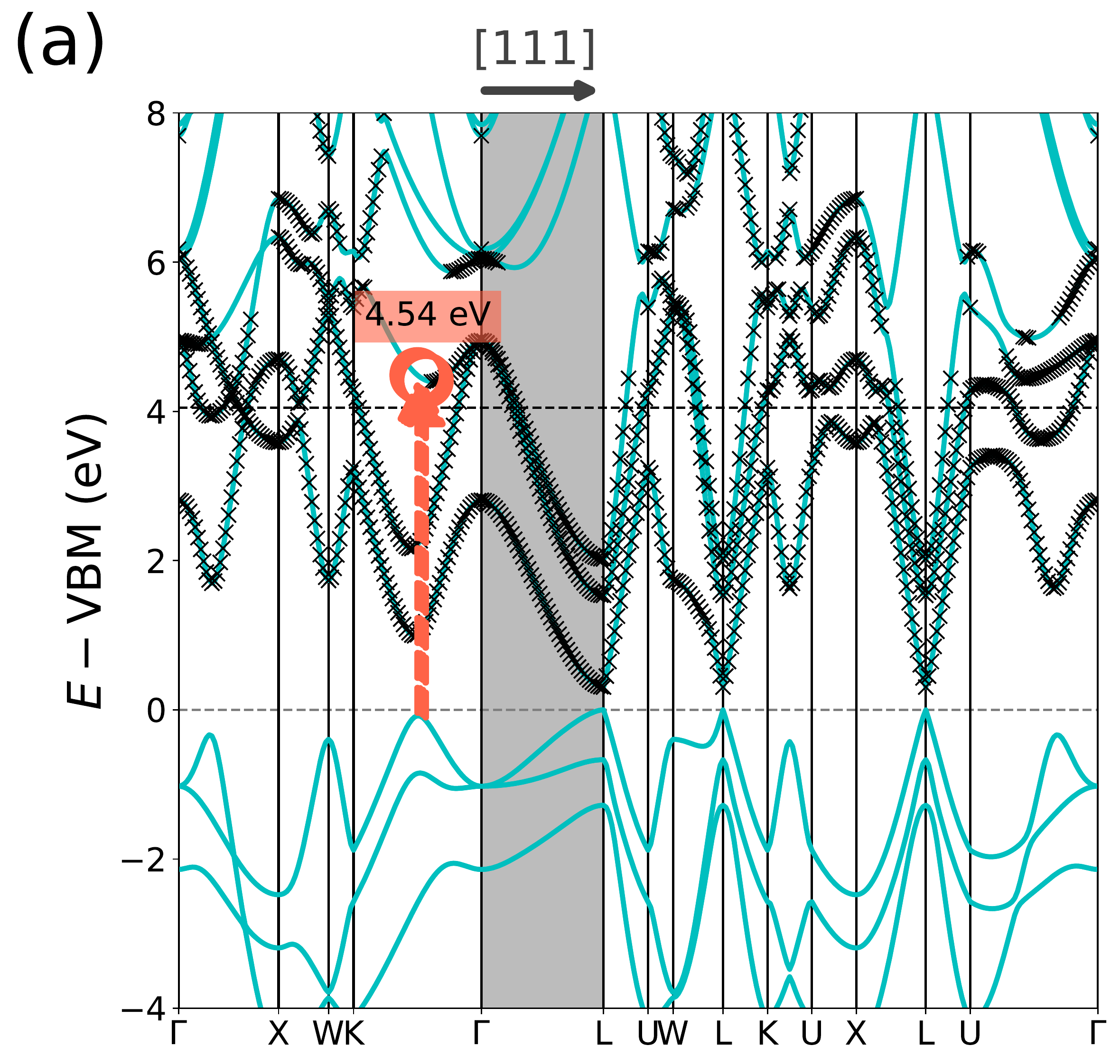}
    \qquad
    \includegraphics[height=0.4\linewidth]{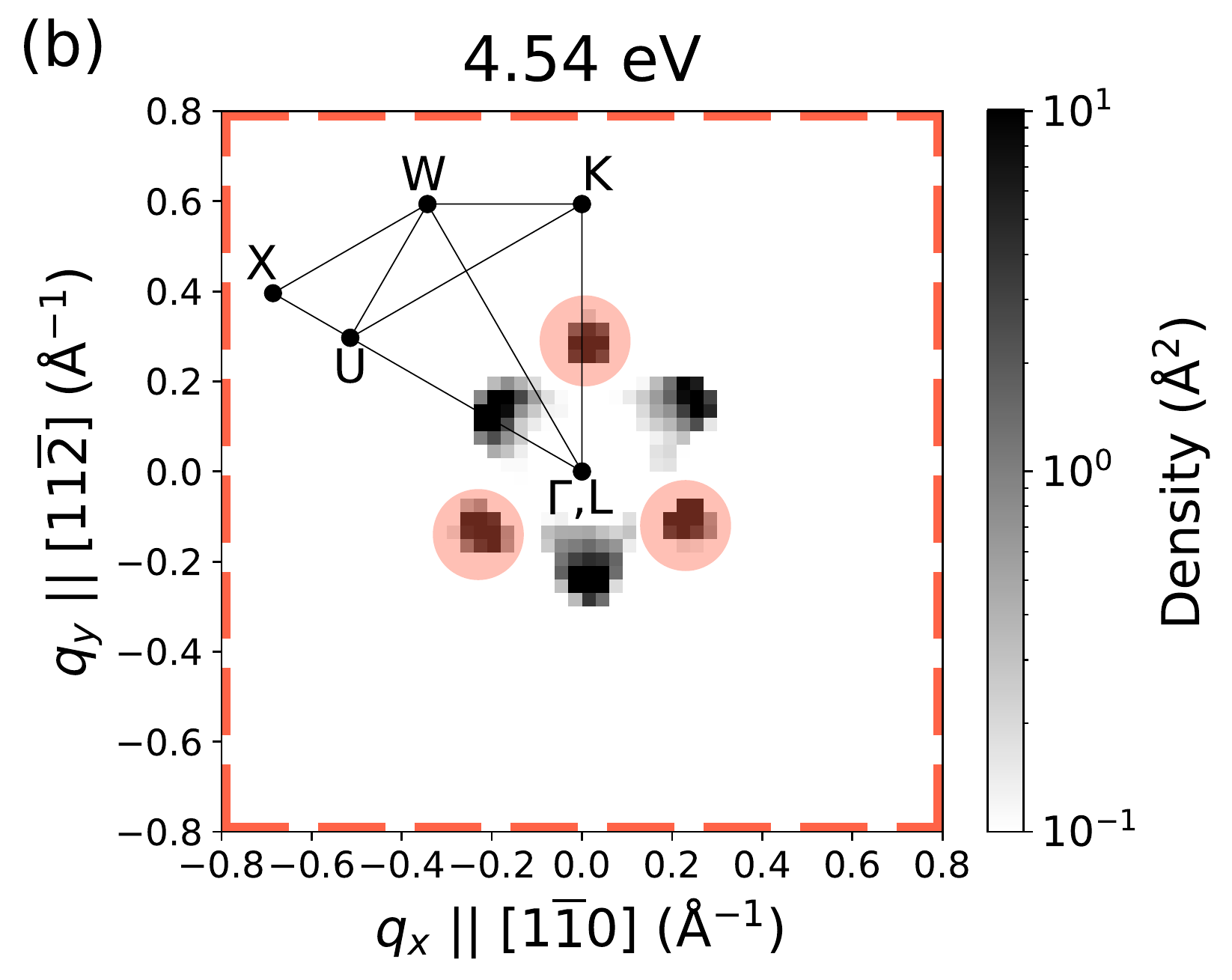}

    \caption{
        (a) Bulk band structure of PbTe: bands (solid curves), high-symmetry
        path $\Gamma$--L parallel to the $[$111$]$ surface normal
        direction (shaded region), valence band maximum (horizontal dashed
        line at 0), vacuum level (horizontal dashed line at 4.05 eV), conduction states with zero surface transmission probabilities ($\times$), direct transitions occurring 
        along high-symmetry paths at the apparent direct threshold of 4.54~eV (vertical dashed arrows), and 
        conduction states corresponding to these direct transitions (circles).
        The arrow-circle pairs are located very close to each other, nearly
        halfway along $\Gamma$--K. 
        (b) Transverse momentum distribution, normalized as probability
        density, including only direct processes at 4.54~eV 
        (square pixels), 
        projection onto the (111) surface plane of high-symmetry paths available in panel (a) (straight lines), 
        and direct transitions depicted in panel (a) (square pixels highlighted by circles).
    }

    \label{fig:bandstruct_trans-mom-distrib_dir}
%\end{figure}
\end{figure*}

For efficient screening, we suggest including at first only direct photoexcitations, 
not only because these excitations can contribute significantly to the MTEs, as in the case of PbTe(111) (Sec.~\ref{subsec:momentotron}), but also because they are significantly less computationally demanding to evaluate. 
When considering only the direct processes, it may be tempting to use publicly-available band structures of prospective photocathode materials. Such band structures, however, generally explore only the high-symmetry paths in the Brillouin zone, which may result in failure to include important contributions from photoexcitation processes occurring at low-symmetry points. 
For example, Fig.~\ref{fig:bandstruct_trans-mom-distrib_dir}(a) shows a bulk band structure that might be found in public databases for PbTe, showing selected paths between the high-symmetry points in the face-centered-cubic Brillouin zone. 
We first eliminate the excited states that have \textit{zero} surface transmission probabilities due to negative perpendicular kinetic energies or group velocities directed away from the surface. The remaining possible vertical transitions then indicate a direct threshold of 4.54~eV. 
In contrast, the actual threshold, which we find by considering all possible transitions in the Brillouin zone, is significantly lower, 4.27~eV (Sec.~\ref{subsec:MTEresult}).

To further illustrate the importance of considering all crystal momenta,
Fig.~\ref{fig:bandstruct_trans-mom-distrib_dir}(b) shows the transverse
momentum distribution at a photon energy of 
4.54~eV, superposed on the projection onto the surface plane of all high-symmetry paths available in the band structure in Fig.~\ref{fig:bandstruct_trans-mom-distrib_dir}(a).
The direct transitions along $\Gamma$--K shown in Fig.~\ref{fig:bandstruct_trans-mom-distrib_dir}(a)
correspond to the cluster of points about halfway along the projected $\Gamma$,L--K path in 
Fig.~\ref{fig:bandstruct_trans-mom-distrib_dir}(b), which has two other copies due to symmetry as indicated in the figure.
There are, however, three additional excitation pockets evident in Fig.~\ref{fig:bandstruct_trans-mom-distrib_dir}(b) 
appearing at points that \textit{do not} correspond to any points in
Fig.~\ref{fig:bandstruct_trans-mom-distrib_dir}(a), which shows no available transitions along any of the paths connecting $\Gamma$ or L to X or U.
This example demonstrates how \textit{all} crystal momenta in the Brillouin zone must be considered to avoid false conclusions about the performance of single-crystal photocathode materials.

%===============================================================
%                 END - COMPUTATIONAL SEARCH
%===============================================================

\section{Conclusions} \label{sec:conclusion}

This work describes, to our knowledge, the first fully \abinit framework for calculating the mean transverse energy (MTE) of 
single-crystal photocathodes. The framework uses the full bulk band structure of the material
under study calculated using density-functional theory. Our framework also considers
various physical processes relevant to photoemission, such as direct
photoexcitation and phonon-mediated photoexcitation, whose transition rates
we calculate from first principles. 
We use our \abinit framework to study the MTE and transverse momentum
distribution from the (111) surface of PbTe as functions of laser photon energy. 
Our results explain the significant discrepancy between the magnitude of the previous MTE predictions\cite{ref:li-sch-p3, ref:li-sch} and the magnitude of our experimentally measured MTEs. Finally, the lessons learned from this case study of PbTe(111) allow us to recommend a computational screening procedure to find low-MTE single-crystal materials based on the photoelectrons that undergo direct photoexcitations.

Despite the good agreement between our calculated MTEs and our experimentally-measured MTEs from PbTe(111), several discrepancies remain.
First, our calculated transverse momentum distributions show the three-fold symmetry of the PbTe(111) crystal surface, whereas our measured distributions show nearly cylindrical symmetry.
We attribute this difference to some combination of
surface reconstructions and relaxations, small rough patches, and atomic steps on the single-crystal surface of PbTe(111). 
Second, our calculated distributions tend to be hollow at the center, whereas our measured distributions tend to be peaked at the center.
This difference is likely due to the contributions of 
the excited electrons that transition directly into vacuum states, which are not considered in this work and whose inclusion would likely improve our agreement with experiment.

We here also consider the effects of temperature, and find results suggesting that 
standard techniques such as lowering the photocathode temperature do not necessarily reduce the MTEs of single-crystal photocathodes.
On the other hand, computational screening of single-crystal materials remains a viable pathway to produce photocathodes with low MTEs. 
In performing the screening, it is important to consider
photoexcitations at \textit{all} crystal momenta instead of only along high-symmetry paths in the Brillouin zone, and to consider both bulk-like transitions and transitions directly into vacuum. 
Finally, once low-MTE candidates are identified, further computational studies using the approach we introduce here should be carried out to determine whether other processes, such as phonon-mediated photoexcitations, significantly affect the MTEs.

%===============================================================
%                   END - CONCLUSION
%===============================================================

\begin{acknowledgements}

This work was supported by the U.S. National Science Foundation under Award PHY-1549132, the Center for Bright Beams (J. K. N., T. A. A., S. K., H. A. P.), 
and by the Director, Office of Science, Office of Basic Energy Sciences of the U.S. Department of Energy, under Contracts No. KC0407-ALSJNT-I0013 and No. DE-AC02-05CH11231 (S. K., H. A. P.).

\end{acknowledgements}

%===============================================================
%                   END - ACKNOWLEDGEMENTS
%===============================================================

\appendix 

\section{Proof that $\{\mathbf{G}_s\} = \{\mathbf{G}_\parallel\}$} \label{sec:app_Gpar}

Let $\mathcal{R}$ be a set containing all vectors $\mathbf{R}$ in a three-dimensional
Bravais lattice and let $\mathcal{G}$ be the set containing all vectors $\mathbf{G}$ in the reciprocal lattice of 
$\mathcal{R}$. By definition, 
\begin{align}
    \mathcal{R} \cdot \mathcal{G} \subset 2 \pi \mathbb{Z},
    \label{eqn:RG}
\end{align}
where $\mathbb{Z}$ is the set of all integers and the dot product is defined as the set containing all possible dot products between the members of $\mathcal{R}$ and the members of $\mathcal{G}$.

For any $\mathcal{R}$ and any surface with unit normal vector $\hat{\mathbf{n}}$, 
\textit{all} lattice vectors along the surface 
form a set $\mathcal{R}_s = \{ \mathbf{R} \in \mathcal{R}: \mathbf{R} \cdot \hat{\mathbf{n}} = 0 \} \subset \mathcal{R}$. 
The set $\mathcal{R}_s$ corresponds to a two-dimensional Bravais lattice because 
for any $\mathbf{R}_{s1}, \mathbf{R}_{s2} \in \mathcal{R}_s$, $\mathbf{R}_{s1} + \mathbf{R}_{s2} \equiv \mathbf{R}_3 \in \mathcal{R}$ and $\mathbf{R}_3 \cdot \hat{\mathbf{n}} = 0$, which imply that $\mathbf{R}_3 \in \mathcal{R}_s$.

Let $\mathcal{G}_\parallel$ be the set of all vectors in $\mathcal{G}$ projected onto the surface defined by the unit normal vector $\hat{\mathbf{n}}$: $\mathcal{G}_\parallel \equiv P \mathcal{G}$, 
where the projection operator $P = 1 - \hat{\mathbf{n}} \hat{\mathbf{n}} \cdot$ applied to the set $\mathcal{G}$ returns the set of the projections of all members of $\mathcal{G}$. 
The set $\mathcal{G}_\parallel$ forms a two-dimensional Bravais lattice because 
for any $\mathbf{G}_{\parallel 1}, \mathbf{G}_{\parallel 2} \in \mathcal{G}_\parallel$, 
there exist $\mathbf{G}_1, \mathbf{G}_2\in \mathcal{G}$ such that 
$\mathbf{G}_{\parallel 1} + \mathbf{G}_{\parallel 2} = P \mathbf{G}_1 + P \mathbf{G}_2 = P (\mathbf{G}_1 + \mathbf{G}_2) = P \mathbf{G}_3 \in \mathcal{G}_\parallel$, where it is clear that $\mathbf{G}_3 \equiv \mathbf{G}_1 + \mathbf{G}_2$ is in $\mathcal{G}$ because $\mathcal{G}$ is a Bravais lattice.

Because $\mathcal{R}_s \subset \mathcal{R}$ it follows from Eq.~\eqref{eqn:RG} that
$\mathcal{R}_s \cdot \mathcal{G} \subset 2 \pi \mathbb{Z}$. Moreover, because $\mathcal{R}_s \cdot \hat{\mathbf{n}} = 0$, the perpendicular components of all members of $\mathcal{G}$ do not affect the dot product values, and thus we can replace $\mathcal{G}$ with $\mathcal{G}_\parallel$ so that
$$
\mathcal{R}_s \cdot \mathcal{G}_\parallel \subset 2 \pi \mathbb{Z}.
$$
This means that each member of $\mathcal{R}_s$ is among those vectors that always equal $2\pi$ times an integer when dotted with any member of $\mathcal{G}_\parallel$. Thus, 
\begin{equation}
    \mathcal{R}_s \subset \mathcal{G}_\parallel^{-1},
    \label{eqn:Rs_in_GparInv}
\end{equation}
where $\mathcal{G}_\parallel^{-1}$ denotes the reciprocal lattice of $\mathcal{G}_\parallel$. 
Note that because $\mathcal{G}_\parallel$ is a two-dimensional Bravais lattice on the surface plane defined by the unit normal vector $\hat{\mathbf{n}}$, $\mathcal{G}_\parallel^{-1}$ is also a two-dimensional Bravais lattice on the same surface plane, and thus $\mathcal{G}_\parallel^{-1} \cdot \hat{\mathbf{n}} = 0$.

We note further that because $\mathcal{G}_\parallel^{-1} \cdot \hat{\mathbf{n}} = 0$, we can replace $\mathcal{G}_\parallel$ with $\mathcal{G}$ in $\mathcal{G}_\parallel^{-1} \cdot \mathcal{G}_\parallel \subset 2 \pi \mathbb{Z}$, yielding
$$
\mathcal{G}_\parallel^{-1} \cdot \mathcal{G} \subset 2 \pi \mathbb{Z},
$$
so that $\mathcal{G}_\parallel^{-1} \subset \mathcal{G}^{-1} = \mathcal{R}$.  Finally, because $\mathcal{G}_\parallel^{-1} \subset \mathcal{R}$ and $\mathcal{G}_\parallel^{-1} \cdot \hat{\mathbf{n}} = 0$, we find
\begin{equation}
  \mathcal{G}_\parallel^{-1} \subset \mathcal{R}_s. 
    \label{eqn:GparInv_in_Rs}  
\end{equation}

Taken together, equations \eqref{eqn:Rs_in_GparInv} and \eqref{eqn:GparInv_in_Rs} now imply that $\mathcal{R}_s = \mathcal{G}_\parallel^{-1}$. Finally, because both $\mathcal{R}_s$ and $\mathcal{G}_\parallel^{-1}$ are Bravais lattices, this also means $\mathcal{R}_s^{-1}= \mathcal{G}_\parallel$, 
so that the set of all $\mathbf{G}_s$ reciprocal to the surface lattice is indeed the same as the set of all $\mathbf{G}_\parallel$, as noted in the main text.

%---------------------------------------------------------------------------------
%---------------------------- END - Appendix Gs = Gpar ---------------------------
%---------------------------------------------------------------------------------

\bibliography{refs} % Use refs.bib

\end{document}